\documentclass[prd,onecolumn,12pt]{revtex4}
\usepackage{amsmath,graphicx}
\usepackage{feynmp}

\newcommand\ba{\begin{eqnarray}}
\newcommand\ea{\end{eqnarray}}
\newcommand\be{\begin{equation}}
\newcommand\ee{\end{equation}}
\newcommand\nn{\nonumber}

\begin{document}

\title{New formulation of $(g-2)_\mu$ hadronic contribution}

\author{Yu.~M.~Bystritskiy}
\email{bystr@theor.jinr.ru}
\affiliation{Joint Institute for Nuclear Research, 141980 Dubna, Russia}

\author{E.~A.~Kuraev}
\email{kuraev@theor.jinr.ru}
\affiliation{Joint Institute for Nuclear Research, 141980 Dubna, Russia}

\author{A.~V.~Bogdan}
\affiliation{Budker Institute of Nuclear Physics, 630090 Novosibirsk, Russia}

\author{F.~V.~Ignatov}
\affiliation{Budker Institute of Nuclear Physics, 630090 Novosibirsk, Russia}

\author{G.~V.~Fedotovich}
\email{G.V.Fedotovich@inp.nsk.su}
\affiliation{Budker Institute of Nuclear Physics, 630090 Novosibirsk, Russia}

\date{\today}% It is always \today, today,

% ------------------------------------------------------------------
\begin{abstract}
In frames of agreement to consider the annihilation of
electron-positron pair to hadrons cross section to be including
the virtual photon polarization effects a new formulation of
hadron contribution to muon anomalous magnetic moment is
suggested. It consists in using the experimentally observed cross
section converted with the known kernels. The lowest order kernel
remains to be the same but some modification of radiative
corrected kernel is needed. The explicit form of this new kernel
is given.
We estimate the accuracy of new formulation on the level
$\delta a^{hadr}_\mu/a^{hadr}_\mu \sim 10^{-5}$.
\end{abstract}

\maketitle

% ------------------------------------------------------------------
\section{Motivation}
% ------------------------------------------------------------------
Anomalous magnetic moment of muon $a_\mu$ is very sensitive laboratory
to search new physics beyond the Standard Model (SM) (see \cite{Nyffeler} and the references therein).
However before driving any premature conclusions about new physics
some caution at the level of precision required of hadronic uncertainties
is SM should be paid \cite{Czarnetski}. The estimation of theoretical and experimental ones becomes
very important.

It is the motivation of our paper to suggest a new, more natural form of
inclusion of hadronic vacuum polarization effects. The theoretical as well the systematic experimental
uncertainties we expect to be considerably reduced.

The SM contributions are usually split into three parts:
$a_\mu = a_\mu^{QED} + a_\mu^{EW} + a_\mu^{hadr}$.
%Here we will not touch $a_\mu^{EW}$.
%We also don't consider the
%hadronic contributions of light-by-light type
The part of $a_\mu^{hadr}$ taking only vacuum polarization effects
(we don't consider hadronic contributions of light-by-light type)
usually is presented in form (see for instance
\cite{krause} and references therein):
\ba
a^{hadr}_\mu = \frac{1}{3}\left(\frac{\alpha}{\pi}\right)^2
\int\limits_{4m_\pi^2}^\infty \frac{ds}{s}
R(s) \left[ K^{(1)}(s) + \frac{\alpha}{\pi} K^{(2)}(s) \right],
\ea
with
\ba
K^{(1)}(s)=\int\limits_0^1
dx\frac{x^2(1-x)}{x^2+\rho(1-x)}, \qquad \rho=\frac{s}{M^2},
\ea
where $M$ is the muon mass and
\be
R(s)=\frac{\sigma^{e^+e^-\to hadr}_0(s)}{\sigma^{e^+e^-\to \mu^+\mu^-}(s)}=12\pi Im_h \Pi(s),
\qquad
\sigma^{e^+e^-\to \mu^+\mu^-}(s) = \frac{4\pi\alpha^2}{3s}.
\ee
The quantity $\sigma^{e^+e^-\to hadr}_0(s)$ which enters the quantity $R(s)$ is
rather unphysical one as well as it does not takes into account the effects of
vacuum polarization of virtual photon ($e^++e^-\to \gamma^* \to hadr$).
The physical one can be obtained by the replacement
\be
Im_h\Pi(s)\to Im_h \left(\frac{\Pi(s)}{1-\Pi(s)}\right)=\frac{Im_h\Pi(s)}{|1-\Pi(s)|^2},
\qquad
\Pi(s)=\Pi_l(s)+\Pi_h(s),
\ee
where $\Pi_l(s)$, $\Pi_h(s)$ are leptonic and hadronic contributions to the vacuum
polarization operator.
Namely the quantity $\sigma^{e^+e^-\to hadr}_{exp.}(s)$, defined as:
\be
\sigma^{e^+e^-\to hadr}_{exp.}(s)=\frac{\sigma^{e^+e^-\to hadr}_0(s)}{|1-\Pi(s)|^2},
\ee
is more relevant to experiment, contrary to Born one $\sigma^{e^+e^-\to hadr}_0(s)$.
In the region of narrow resonances the application of this formula must be performed with
some care \cite{FedotovichSolodov}.

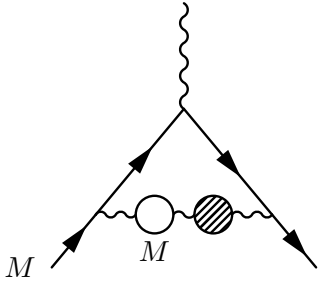
\begin{figure}
\begin{center}
\begin{fmffile}{Fig1}
\begin{fmfgraph*}(100,200)
\fmfleft{mu1}
\fmfright{mu2}
\fmftop{gamma}
%\fmfdot{v1,v2,v2}
\fmf{fermion,tension=2}{mu1,v1}
\fmf{fermion}{v1,v3,v2}
\fmf{fermion,tension=2}{v2,mu2}
\fmf{photon,tension=2}{v3,gamma}
\fmffreeze
\fmfv{decor.shape=circle,decor.filled=0,label.angle=-90,label.dist=10}{lepton_contribution}
\fmfv{decor.shape=circle,decor.filled=.5}{hadron_contribution}
\fmf{photon}{v1,lepton_contribution}
\fmf{photon}{lepton_contribution,hadron_contribution}
\fmf{photon}{hadron_contribution,v2}
\fmflabel{$M$}{mu1}
\fmflabel{$M$}{lepton_contribution}
\end{fmfgraph*}
\end{fmffile}
\end{center}
\vspace{-4cm}
\caption{Subtracting Feynman diagram.}
\label{SubtractingFD}
\end{figure}
%

% ------------------------------------------------------------------
\section{Second order kernel modification}
% ------------------------------------------------------------------
Keeping this definition in mind one must revise the formulae for $a^{hadr}_\mu$,
cited above.
Really, one must replace in integrands of $a^{hadr}_\mu$: $\sigma^{e^+e^-\to hadr}_0(s)\to
\sigma^{e^+e^-\to hadr}_{exp.}(s)$.
Kernel $K^{(1)}(s)$ remains the same, but the kernel $K^{(2)}(s)$ must be
modified to avoid the double counting. The modification consists in
eliminating of contributions of all Feynman diagrams containing two
polarization of
vacuum insertions (both hadronic, leptonic sort and the mixed ones).
It results in omitting the contributions of $K^{(2b,2c)}(s)$ in terminology
of ref. \cite{krause}. As for $K^{(2a)}(s)$ it must be modified, in such a
way to extract the contribution of
such Feynman diagram (see Fig.~\ref{SubtractingFD}) which contains
polarization operator for muon case with hadronic one. So our result
consists in replacement $K^{(2a)}(s)$ \cite{Barbieri:1974nc}:
\ba K^{(2a)}(s) &=& 2\left\{-\frac{139}{144} + \frac{115}{72}\rho
+\left(\frac{19}{12}-\frac{7}{36}\rho + \frac{23}{144}\rho^2+\frac{1}{\rho-4}\right) L+
\right.\nn\\
&+&\frac{1}{\Delta}\left(-\frac{4}{3}+\frac{127}{36}\rho - \frac{115}{72}\rho^2 + \frac{23}{144}\rho^3\right)
\ln y +
\left(\frac{9}{4}+\frac{5}{24}\rho - \frac{1}{2}\rho^2 - \frac{2}{\rho}\right) \xi_2 + \nn\\
&+&\frac{5}{96}\rho^2 L^2 +
\frac{1}{\Delta}\left(-\frac{1}{2}\rho + \frac{17}{24}\rho^2 - \frac{7}{48}\rho^3\right)
L \ln y + \nn\\
&+&
\left(\frac{19}{24}+\frac{53}{48}\rho - \frac{29}{96}\rho^2 - \frac{1}{3\rho} + \frac{2}{\rho-4}\right)
\ln^2 y + \nn\\
&+&
\frac{1}{\Delta}\left(-2\rho + \frac{17}{6}\rho^2 - \frac{7}{12}\rho^3\right)
D_p(\rho) + \nn\\
&+& \left.\frac{1}{\Delta} \left(\frac{13}{3}-\frac{7}{6}\rho +
\frac{1}{4}\rho^2 - \frac{1}{6}\rho^3 - \frac{4}{\rho-4}\right)
D_m(\rho) + \left(\frac{1}{2}-\frac{7}{6}\rho +
\frac{1}{2}\rho^2\right)T(\rho)\right\},
\ea
with $L=\ln(s/M^2)$, $\Delta=\sqrt{\rho(\rho-4)}$, $\xi_2=\pi^2/6$ and
\ba
y &=& \frac{\sqrt{\rho} - \sqrt{\rho-4}}{\sqrt{\rho} + \sqrt{\rho-4}}, \nn\\
D_p(\rho) &=& Li_2(y) + \ln y ~ \ln(1-y) - \frac{1}{4} \ln^2y - \xi_2, \nn\\
D_m(\rho) &=& Li_2(-y) + \frac{1}{4} \ln^2y + \frac{1}{2} \xi_2, \nn\\
T(\rho) &=& -6 Li_3(y) - 3 Li_3(-y) + \ln^2y~\ln(1-y) + \frac{1}{2} (\ln^2y + 6 \xi_2)\ln(1+y) + \nn\\
&+& 2 \ln y (Li_2(-y) + 2 Li_2(y)), \nn
\ea
\ba
Li_2(y) = -\int\limits^y_0 \frac{dx}{x} \ln(1-x), \qquad
Li_3(y) =  \int\limits^y_0 \frac{dx}{x} Li_2(x),
\ea
%
%\ba
%K^{(2a)}(s)=2\frac{1}{\rho}\left[a_1+b_1L+\frac{1}{\rho}(a_2+b_2L+c_2L^2)+
%\frac{1}{\rho^2}(a_3+b_3L+c_3L^2)+
%\frac{1}{\rho^3}(a_4+b_4L+c_4L^2)\right],
%\ea
%%
%\be
%\begin{array}{lll}
%a_1=\frac{223}{54}-2\xi, & b_1=-\frac{23}{36}, & \\
%a_2=\frac{8785}{1152}-\frac{37}{8}\xi, & b_2=-\frac{367}{216}, & c_2=\frac{19}{144}, \\
%a_3=\frac{13072841}{432000}-\frac{883}{40}\xi, & b_3=-\frac{10079}{3600}, & c_3=\frac{141}{80}, \\
%a_4=\frac{2034703}{16000}-\frac{3903}{40}\xi, & b_4=-\frac{6517}{1800}, & c_4=\frac{961}{80}, \\
%\end{array}
%\ee
%
by the new one:
\be
 \bar{K}^{(2)}(s) = K^{(2a)}(s)-\left.K^{(2b)}(s)\right|_{m_f=M},
 \label{modifyedKernel}
\ee
with
\ba
K^{(2b)}(s)_{m_f=M} &=&2\int\limits_0^1dx\frac{x^2(1-x)}{x^2+\rho(1-x)}\Pi(1,x), \nonumber \\
\Pi(1,x)&=&-\frac{8}{9}+\frac{b^2}{3}-b\left(\frac{1}{2}-\frac{b^2}{6}\right)\ln\frac{b-1}{b+1}, \nonumber \\
b=\frac{2-x}{x}.
\ea
The quantity $K^{(2b)}(s)_{m_f=M}$ can be calculated analytically:
\ba
K^{(2b)}(s)_{m_f=M}=
 \frac{2}{\rho}\left[\frac{8}{9}\rho^2+\frac{35}{36}\rho-\frac{4}{3}\xi_2
-\frac{1}{\Delta}[L_-P_1(x_-)-L_+P_1(x_+)]-\right.&&\nonumber \\
-\left.\frac{1}{\Delta}[Li_-P_2(x_-)-Li_+P_2(x_+)]\right],&&
\ea
with $x_\pm=(\rho\pm\Delta)/2$, $\Delta=\sqrt{\rho(\rho-4)}$ and
\ba
L_\pm=\ln\frac{x_\pm}{x_\pm-1}, &\qquad& Li_\pm=Li_2(1-x_\mp), \nonumber \\
P_1(z)=-\frac{5}{9}z^4-\frac{4}{3}z^3+\frac{4}{3}z^2, &\qquad&
P_2(z)=\frac{1}{3}z^4-2z^2+\frac{4}{3}z.
\ea
For expansion into series by powers of $\rho^{-1}$ we have:
\ba
\bar{K}^{(2)}(s)&=&2\frac{1}{\rho}\left[\bar{a}_1+\bar{b}_1L+\frac{1}{\rho}(\bar{a}_2+\bar{b}_2L+\bar{c}_2L^2)+
\frac{1}{\rho^2}(\bar{a}_3+\bar{b}_3 L+\bar{c}_3L^2)+ \right.\nn\\
&+& \left.\frac{1}{\rho^3}(\bar{a}_4+\bar{b}_4L+\bar{c}_4L^2)
+\frac{1}{\rho^4}(\bar{a}_5+\bar{b}_5 L+\bar{c}_5 L^2) \right] + O(\rho^{-6}),
\label{K2Expansion}
\ea
with
\be
\begin{array}{lll}
\bar{a}_1=\frac{50}{27}-\frac{2}{3}\xi_2, & \bar{b}_1=-\frac{23}{36}, & \\
\bar{a}_2=\frac{9241}{1152}-\frac{103}{24}\xi_2, & \bar{b}_2=-\frac{487}{216}, & \bar{c}_2=\frac{43}{144}, \\
\bar{a}_3=\frac{15256601}{432000}-\frac{803}{40}\xi_2, & \bar{b}_3=-\frac{29279}{3600}, & \bar{c}_3=\frac{221}{80}, \\
\bar{a}_4=\frac{66452261}{432000}-\frac{10829}{120}\xi_2, & \bar{b}_4=-\frac{57917}{1800}, & \bar{c}_4=\frac{3763}{240}, \\
\bar{a}_5=\frac{18433084459}{27783000}-\frac{13877}{35}\xi_2, & \bar{b}_5=-\frac{34443349}{264600}, & \bar{c}_5=\frac{47651}{630}.
\end{array}
\ee
So our final result have a form for hadronic contribution to anomalous magnetic
moment of muon is:
\ba
a^{hadr}_\mu = \frac{1}{3}\left(\frac{\alpha}{\pi}\right)^2
\int\limits_{4m_\pi^2}^\infty \frac{ds}{s}
R^h_{exp.}(s) \left[ K^{(1)}(s) + \frac{\alpha}{\pi} \bar{K}^{(2)}(s) \right],
\label{result}
\ea
where
$R^h_{exp.}(s) = \sigma^{e^+e^-\to hadr}_{exp.}(s)/\sigma^{e^+e^-\to \mu^+\mu^-}(s)$ and $\bar{K}^{(2)}(s)$
is given above (see (\ref{modifyedKernel}), (\ref{K2Expansion})).

% ------------------------------------------------------------------
\section{Discussion}
\begin{figure}
\begin{center}
\begin{fmffile}{Fig2}
\begin{fmfgraph*}(100,200)
\fmfleft{mu1}
\fmfright{mu2}
\fmftop{gamma}
%\fmfdot{v1,v2,v2}
\fmf{fermion,tension=2}{mu1,v1}
\fmf{fermion}{v1,v4,v3,v5,v2}
\fmf{fermion,tension=2}{v2,mu2}
\fmf{photon,tension=2}{v3,gamma}
\fmffreeze
\fmfv{decor.shape=circle,decor.filled=0,label.angle=-90,label.dist=10}{lepton_contribution}
\fmfv{decor.shape=circle,decor.filled=.5}{hadron_contribution}
\fmf{photon}{v1,hadron_contribution}
\fmf{photon}{hadron_contribution,v2}
\fmf{photon}{v4,lepton_contribution}
\fmf{photon}{lepton_contribution,v5}
\end{fmfgraph*}
\end{fmffile}
\end{center}
\vspace{-4cm}
\caption{Typical contribution with lepton and hadron vacuum polarization associated with
different virtual photon lines enhanced by logarithmical factor.}
\label{TypicalHigherOrderFD}
\end{figure}
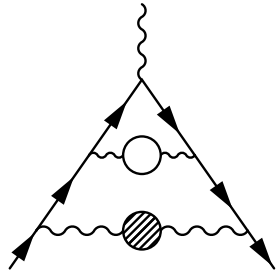

% ------------------------------------------------------------------
The set of FDs contribution with lepton and hadron vacuum polarization associated with
different virtual photon lines cannot be considered with the method discussed above.
Their contribution (see Fig.~\ref{TypicalHigherOrderFD}) enhanced
by logarithmical factor can be estimated as $\delta a^{hadr}_\mu \sim
(\alpha/\pi)^2(1/3)\ln M^2/m_e^2 \approx 2 \cdot10^{-5} a^{hadr}_\mu$.
Fortunately this is beyond the modern experimental possibilities.

%The manifestation of hadronic effects also takes place in leptonic imaginary part of
%vacuum polarization $Im_l(\Pi(s))$. It can be written in the form similar the one given above
%(\ref{result}) with $R^h_{exp.}(s) \to R^l_{exp.}(s) = \sigma^{e^+e^-\to leptons}_{exp.}(s)/\sigma^{e^+e^-\to \mu^+\mu^-}(s)$.
%This replacement however results in double counting of pure electrodynamic Feynman diagrams contributions.
%Namely all the QED diagrams with leptonic vacuum polarization must be eliminated from $a_\mu^{QED}$.
%There is explicit expression for this contributions in literature \cite{Barbieri:1974nc,Lautrup}.
%In this way however we encounter another problem of experimental measuring
%$\sigma^{e^+e^-\to leptons}_{exp.}(s)$.

%Another way to take into account the hadronic modification of lepton contribution
%consists in direct calculation of mixed Feynman diagrams. We did not touch this problem here.
%
%

% ------------------------------------------------------------------
\begin{acknowledgements}
% ------------------------------------------------------------------
We are grateful to E.~Solodov, S.~Eidelman, B.~Khazin and other participants of
Budker Institute of Nuclear Physics (Novosibirsk) seminar.
This work is supported by RFBR grants 03-02-17-077.
\end{acknowledgements}

\end{document}